\documentclass[10pt,twoside]{article}
\usepackage[applemac]{inputenc}
\usepackage{amssymb,amsmath}
\def\volnum{Volume 37 no 1, 2012}

\usepackage{aflbcours}
\pdebut{1}
\def\pacs#1{\LP P.A.C.S.: #1}
\title{On the Solvability of Maxwell's Equations}
\titleshort{Solvability \dots}
\author{Wolfgang Engelhardt\footnotemark[1]}
\footnotetext[1]{Home address: Fasaneriestrasse 8, D-80636 M\"unchen, Germany 
\\ 
\indent\hspace{1.4mm} 
E-mail address: wolfgangw.engelhardt@t-online.de }
\authorshort{Engelhardt}
\address{retired from: \\ Max-Planck-Institut f\"ur Plasmaphysik, D-85741 
Garching, Germany}

\begin{document}
\maketitle

\vskip 1cm
\begin{abstract}
R\'ESUM\'E. Comme compl\'ement \`a une \'etude publi\'ee dans ce journal en 2005, nous pr\'esentons des calculs explicites des champs \'electro-magn\'etiques selon la th\'eorie de Maxwell moyennant autant la jauge de Lorenz que celle de Coulomb. On peut obtenir des expressions analytiques quand la source des champs est un dip\^ole \'electrique oscillant. Comme avant, on trouve que les champs, qui sont calcul\'es selon des m\'ethodes diff\'erentes, sont contradictoires. En outre, la raison pour le d\'esaccord est d\'ecouverte: Contrairement \`a l'expectative les int\'egrales retard\'ees ne satisfont pas les \'equations non homog\`enes d'onde qu'elles sont cens\'ees r\'esoudre.

{\it 
ABSTRACT. Complementing a study which was published in this journal in 2005 we present explicit calculations of fields predicted by Maxwell's equations both in Lorenz and in Coulomb gauge. Analytic expressions are obtainable, when the source of the fields is an oscillating electric dipole. As before it is found that the fields calculated by different methods are at variance. In addition, the reason for the discrepancies is revealed: The retarded integrals turn out not to satisfy the inhomogeneous wave equations which they are supposed to solve.}
\end{abstract}

\section{Introduction}

In papers by Vladimir Onoochin [1] and by the present author [2] it was shown that the solutions of Maxwell's equations for point charges depend on the chosen gauge. Whereas in Lorenz gauge ($c\,\,\mbox{div}\,\vec {A}+{\partial \phi } \mathord{\left/ {\vphantom {{\partial \phi } {\partial t=0}}} \right. \kern-\nulldelimiterspace} {\partial t=0})$ one obtains the well known Li\'{e}nard-Wiechert fields, a calculation in Coulomb gauge ($\mbox{div}\,\vec {A}=0)$ leads to indefinite or diverging integrals for the electromagnetic fields. The reason for this discrepancy was not entirely clarified, but in [2] it was suspected that an internal inconsistency of Maxwell's equations is responsible for the ambiguity in the solutions. In 1971 Donald Dunn [3] pointed out that Maxwell's differential equations yield satisfactory results for the quasi-static instantaneous fields, but inclusion of the displacement current resulted in certain discrepancies which were, however, not resolved in Dunn's book.

Our analysis in [2] led us to the suspicion that the mixture of elliptic and hyperbolic equations, which constitute Maxwell's system, actually causes the incompatibilities between instantaneous solutions derived from elliptic equations (Coulomb field, e.g.) and the travelling wave fields as derived from hyperbolic equations. A combination of both results in inhomogeneous wave equations such as $c^2\Delta \phi -{\partial ^2\phi } \mathord{\left/ {\vphantom {{\partial ^2\phi } {\partial t^2}}} \right. \kern-\nulldelimiterspace} {\partial t^2}=f\left( {\vec {x},\,t} \right)$ which have a strange property: The time $t$ appearing in the d'Alembert operator is the same time  $t$ which describes the temporal behaviour of the source. The observation point and the source may be separated by many light-years, but in the differential equation the events of emission and detection are connected as if they were happening at the same time $_{ }t$. Of course, in the ``retarded'' solutions one dates back the time $t'_{,}$ at which the source is evaluated, by the travel time of the wave: $t'=t-r \mathord{\left/ {\vphantom {r c}} \right. \kern-\nulldelimiterspace} c$. In view of this disparate conception of time in the differential equation and in the solution it appears, however, questionable whether the production of waves is correctly modelled by the inhomogeneous wave equations which follow from Maxwell's first order system.

Whereas in [2] we concentrated on the production of fields by single moving point charges, we use in this paper an oscillating dipole of vanishing extension as the source for the fields. This method is due to Hertz who calculated his famous radiation dipole field on the same basis. The advantage is not only that one has analytical solutions for potentials and fields in Lorenz gauge, but one can also use the dipole source for the calculation of the fields in Coulomb gauge. The pertaining improper integrals do not diverge, but have well defined limiting values. 

After a recapitulation of the standard method of solution in Sect. 2, we introduce in Sect. 3 the Hertzian solution for a point-like dipole in Lorenz gauge and compare it with the solution calculated in Coulomb gauge. As expected from the results in [2] the fields obtained by the two methods are indeed different. In Section 4 we analyze a special inhomogeneous wave equation with an extended source. Surprisingly it turns out that the standard retarded solution does not satisfy the given wave equation. This explains then why we have found so many inconsistencies in [2]. The tacit assumption that Duhamel's principle could be used to construct a solution of the inhomogeneous equation on the basis of solutions of the homogeneous equation turns out to be wrong in case of wave equations. The method is only suitable for the diffusion equation with its intrinsic infinite propagation velocity that connects temperature, for example, instantly with the heat source at any time. Finally, in Section 4 we discuss some consequences of our findings.

\vskip 30pt
\section{Maxwell's equations solved for point charges in Lorenz gauge}
In vacuo Maxwell's first order equations are formulated as follows [4]:
\begin{equation}
\label{eq1}
\mbox{div}\,\vec {E}=4\pi \,\rho 
\end{equation}
\begin{equation}
\label{eq2}
c\,\,\mbox{rot}\,\vec {E}=-\frac{\partial \vec {B}}{\partial t}
\end{equation}
\begin{equation}
\label{eq3}
\mbox{div}\,\vec {B}=0
\end{equation}
\begin{equation}
\label{eq4}
c\,\,\mbox{rot}\,\vec {B}=4\pi \,\vec {j}+\frac{\partial \vec {E}}{\partial t}
\end{equation}
The potential ansatz
\begin{equation}
\label{eq5}
\vec {E}=-\frac{1}{c}\frac{\partial \vec {A}}{\partial t}-\nabla \phi \;,\quad \vec {B}=\mbox{rot}\,\vec {A}
\end{equation}
and the Lorenz condition 
\begin{equation}
\label{eq6}
\mbox{div}\,\vec {A}=-\frac{1}{c}\frac{\partial \phi }{\partial t}
\end{equation}
yield with (1 -- 4) two second order wave equations of the same structure:
\begin{equation}
\label{eq7}
\Delta \phi -\frac{1}{c^2}\frac{\partial ^2\phi }{\partial t^2}=-4\pi \,\rho 
\end{equation}
\begin{equation}
\label{eq8}
\Delta \vec {A}-\frac{1}{c^2}\frac{\partial ^2\vec {A}}{\partial t^2}=-\frac{4\pi }{c}\,\vec {j}
\end{equation}
Li\'{e}nard and Wiechert have solved them for point charges adopting 
\begin{equation}
\label{eq9}
\vec {j}=\rho \,\vec {v}\left( t \right)\; , \quad e=\int\!\!\!\int\!\!\!\int {\rho \left( {\vec {x}{\kern 1pt},\,t} \right)} \,d^3x
\end{equation}
and obtained:
\begin{equation}
\label{eq10}
\phi \left( {\vec {x},\,t} \right)=\left[ {\frac{e}{R-{\vec {R}\cdot \vec {v}} \mathord{\left/ {\vphantom {{\vec {R}\cdot \vec {v}} c}} \right. \kern-\nulldelimiterspace} c}} \right]_{t'=t-R \mathord{\left/ {\vphantom {R c}} \right. \kern-\nulldelimiterspace} c} 
\end{equation}
\begin{equation}
\label{eq11}
\vec {A}\left( {\vec {x},\,t} \right)=\left[ {\frac{e\,\vec {v}}{c\,R-\vec {R}\cdot \vec {v}}} \right]_{t'=t-R \mathord{\left/ {\vphantom {R c}} \right. \kern-\nulldelimiterspace} c} 
\end{equation}
where the vector $\vec {R}$ denotes the distance between the observation point and the position of the charge at the retarded time $t'=t-R \mathord{\left/ {\vphantom {R c}} \right. \kern-\nulldelimiterspace} c$:
\begin{equation}
\label{eq12}
\vec {R}=\vec {x}-\vec {x}\,'\left( {t'} \right)
\end{equation}
With (\ref{eq5}) the fields may be derived from the retarded potentials (\ref{eq10}) and (\ref{eq11}):
\begin{eqnarray}
\label{eq13}
 \vec {E}\left( {\vec {x}{\kern 1pt},\,t} \right)=\frac{e}{\lambda ^3}\,\left( {\frac{\vec {R}}{R^3}-\frac{\vec {v}}{c\,R^2}} \right)\,\left( {1-\frac{v^2}{c^2}+\frac{1}{c^2}\,\vec {R}\cdot \frac{\,d{\kern 1pt}\vec {v}}{dt'}} \right)-\frac{e}{c^2R\,\lambda ^2}\frac{\,d{\kern 1pt}\vec {v}}{dt'} \nonumber \\ 
 \vec {B}\left( {\vec {x}{\kern 1pt},\,t} \right)=\frac{\vec {R}}{R}\times \vec {E}\left( {\vec {x}{\kern 1pt},\,t} \right)\;,\quad \lambda =1-\frac{\vec {R}\cdot \vec {v}\left( {t'} \right)}{R\,c }\quad \quad \quad \quad \quad 
\end{eqnarray}

Hertz has investigated the particular case of an oscillating electric dipole with moment $e\,\vec {d}=\vec {p}\left( t \right)$ located at the origin. If the amplitude is small compared to the wavelength of the emitted radiation and small compared to the distance $r$ at which the fields are observed, one may replace (\ref{eq11}) by the simplified expression [5]:
\begin{equation}
\label{eq14}
\vec {A}\left( {\vec {x}{\kern 1pt},\,t} \right)=\frac{\dot {\vec {p}}\left( {t-r \mathord{\left/ {\vphantom {r c}} \right. \kern-\nulldelimiterspace} c} \right)}{c\,r}\;,\quad r= \sqrt {x^2+y^2+z^2}
\end{equation}
and obtain with the Lorenz condition (\ref{eq6}) for the scalar potential instead of (\ref{eq10}):
\begin{equation}
\label{eq15}
\phi \left( {\vec {x}{\kern 1pt},\,t} \right)=-\,\mbox{div}\frac{\vec {p}\left( {t-r \mathord{\left/ {\vphantom {r c}} \right. \kern-\nulldelimiterspace} c} \right)}{r}=\frac{\vec {p}\cdot \vec {r}}{r^3}+\frac{\dot {\vec {p}}\cdot \vec {r}}{c\,r^2}
\end{equation}
Insertion into the potential ansatz (\ref{eq5}) yields for the fields of Hertz's dipole:
\begin{eqnarray}
\label{eq16}
\vec {E}\left( {\vec {x}{\kern 1pt},\,t} \right)=-\frac{\vec {p}}{r^3}+\frac{3\left( {\vec {p}\cdot \vec {r}} \right)\,\vec {r}}{r^5}-\frac{\dot {\vec {p}}}{c\,r^2}+\frac{3\left( {\dot {\vec {p}}\cdot \vec {r}} \right)\,\vec {r}}{c\,r^4}-\frac{\ddot {\vec {p}}}{c^2\,r}+\frac{\left( {\ddot {\vec {p}}\cdot \vec {r}} \right)\,\vec {r}}{c^2\,r^3} \nonumber \\ 
 \vec {B}\left( {\vec {x}{\kern 1pt},\,t} \right)=\frac{\dot {\vec {p}}\times \vec {r}}{c\,r^3}+\frac{\ddot {\vec {p}}\times \vec {r}}{c^2\,r^2} \quad \quad \quad \quad \quad \quad \quad \quad 
\end{eqnarray}

It is interesting to note that both (10, 11) and (14, 15) are actually particular solutions of the homogeneous wave equations (\ref{eq7}) and (\ref{eq8}) with vanishing sources $\rho \,\,\mbox{and}\,\vec {j}$. The Laplace operator applied on the potential of a point charge is locally not defined at the charge's position so that the infinite density of the charge cloud does not show up in the differential equation. 

\vskip 30pt
\section{Maxwell's equations solved for point-like dipoles in Coulomb gauge}
In Sect. 3 of Ref. [2] it was shown that a solution of (1 -- 5) in Coulomb gauge
\begin{equation}
\label{eq17}
\mbox{div}\,\vec {A}=0
\end{equation}
was not possible, since the improper integrals to be evaluated were either undetermined or would even diverge. One could suspect that this impasse was caused by the assumption of point charges which introduce singularities in the fields. If one uses point-like dipoles, however, instead of single charges, it turns out that the integrals in question do converge. This way a direct comparison of the fields calculated in both gauges (\ref{eq6}) and (\ref{eq17}) can be made. 

With (\ref{eq17}) and (\ref{eq5}) equation (\ref{eq1}) reduces to the Poisson equation 
\begin{equation}
\label{eq18}
\Delta \phi _C =-4\pi \,\rho 
\end{equation}
for the instantaneous Coulomb potential which has for a dipole the solution
\begin{equation}
\label{eq19}
\phi _C \left( {\vec {x}{\kern 1pt},\,t} \right)=\frac{\vec {p}\left( t \right)\cdot \vec {r}}{r^3}
\end{equation}
Equation (\ref{eq8}) becomes
\begin{equation}
\label{eq20}
\Delta \vec {A}_C -\frac{1}{c^2}\frac{\partial ^2\vec {A}_C }{\partial t^2}=-\frac{4\pi }{c}\,\vec {j}+\frac{1}{c}\nabla \frac{\partial \phi _C }{\partial t}
\end{equation}
The solution may be split into two terms $\vec {A}_C =\vec {A}_1 +\vec {A}_2 $ where the first one is identical with the Lorenz vector potential (\ref{eq14})
\begin{equation}
\label{eq21}
\vec {A}_1 \left( {\vec {x}{\kern 1pt},\,t} \right)=\frac{\dot {\vec {p}}\left( {t-r \mathord{\left/ {\vphantom {r c}} \right. \kern-\nulldelimiterspace} c} \right)}{c\,r}
\end{equation}
and the second one obeys the inhomogeneous wave equation
\begin{equation}
\label{eq22}
\Delta \vec {A}_2 -\frac{1}{c^2}\frac{\partial ^2\vec {A}_2 }{\partial t^2}=\frac{1}{c}\nabla \frac{\partial \phi _C }{\partial t}=\frac{\dot {\vec {p}}\left( t \right)}{c\,r^3}-\frac{3\left( {\dot {\vec {p}}\left( t \right)\cdot \vec {r}} \right)\,\vec {r}}{c\,r^5}
\end{equation}
We express the solution as a retarded integral
\begin{equation}
\label{eq23}
\vec {A}_2 \left( {\vec {x}{\kern 1pt},\,t} \right)=\frac{-1}{4\pi \,c}\int\!\!\!\int\!\!\!\int {\left( {\frac{\dot {\vec {p}}\left( {t-{r'} \mathord{\left/ {\vphantom {{r'} c}} \right. \kern-\nulldelimiterspace} c} \right)}{r'^3}-\frac{3\left( {\dot {\vec {p}}\left( {t-{r'} \mathord{\left/ {\vphantom {{r'} c}} \right. \kern-\nulldelimiterspace} c} \right)\cdot \vec {r}\,'} \right)\,\vec {r}\,'}{r'^5}} \right)} \frac{d^3r'}{\left| {\vec {x}-\vec {r}\,'} \right|}
\end{equation}
in analogy to the method of solution which was applied to obtain (10, 11) from (7, 8). Let us assume that the dipole is oriented in $z$--direction. Because of the axial symmetry of the problem it is sufficient to evaluate the $z$--component of $\vec {A}_2 $ at $y=0$:
\begin{eqnarray}
\label{eq24}
A_{2z} =\frac{1}{4\pi \,c}\int\limits_0^\infty {dr'} \int\limits_0^\pi {d\theta '\int\limits_0^{2\pi } {d\varphi '\frac{\left( {3\,\cos ^2\theta '-1} \right) \dot {p}\left( {t-r'/c} \right)\sin \theta}{r'R }} } \nonumber\\
R=\sqrt {r^2+r'^2-2\,r\,r'\left( \cos \varphi '\sin \theta '\sin \theta +\cos \theta '\cos \theta \right)}\quad\quad
\end{eqnarray}
where we have used spherical coordinates in the integrand:
$\newline x=r\sin \theta \cos \varphi \,,\;y=r\sin \theta \sin \varphi \,,\;z=r\cos \theta .$ For the $x$--component of $\vec {A}_2 $ we obtain at $y=0$:
\begin{equation}
\label{eq25}
A_{2x} =\frac{1}{4\pi \,c}\int\limits_0^\infty {dr'} \int\limits_0^\pi {d\theta '\int\limits_0^{2\pi } {d\varphi '\frac{3\,\dot {p}\left( {t-r'/c} \right)\,\cos \varphi '\cos \theta '\sin ^2\theta '\,}{r'R} } } 
\end{equation}
Carrying out the integration over the angles we find:
\begin{eqnarray}
\label{eq26}
A_{2z} = \int\limits_0^r {\frac{r'\, g \,}{\,5\,c\,r^3}\,dr'
+\int\limits_r^\infty {\frac{r^2 g\,\,}{5\,c\,r'^4}\,dr'} } \,,\quad g=\dot {p}\left( {t-{r'} \mathord{\left/ {\vphantom {{r'} c}} \right. \kern-\nulldelimiterspace} c} \right)\left( {2-3\sin ^2\theta } \right)\quad\quad\nonumber \\ 
A_{2x} =\int\limits_0^r {\frac{3\,r'\,\sin 2\theta \,\dot {p}\left( {t-{r'} \mathord{\left/ {\vphantom {{r'} c}} \right. \kern-\nulldelimiterspace} c} \right)}{10\,c\,r^3}\,dr'+\int\limits_r^\infty {\frac{3\,r^2\,\sin 2\theta \,\dot {p}\left( {t-{r'} \mathord{\left/ {\vphantom {{r'} c}} \right. \kern-\nulldelimiterspace} c} \right)}{10\,c\,r'^4}\,dr'} }\quad\quad\;\;
\end{eqnarray}
In order to carry out the remaining radial integration we choose:
\begin{equation}
\label{eq27}
p\left( {t-r \mathord{\left/ {\vphantom {r c}} \right. \kern-\nulldelimiterspace} c} \right)=\,\sin\omega \,\left( {t-r \mathord{\left/ {\vphantom {r c}} \right. \kern-\nulldelimiterspace} c} \right)
\end{equation}
The integrals in (\ref{eq26}) can now be evaluated yielding in Cartesian coordinates with $\omega =c\,k$:
\begin{eqnarray}
\label{eq28}
\vec {A}_2 \left( {\vec {x}{\kern 1pt},\,t} \right)=\frac{f\left( {r{\kern 1pt},\,\,t} \right)}{30\,c\,k}\,\nabla \left( {\frac{\vec {p}}{p}\cdot \nabla \frac{1}{r}} \right)\;,\quad r=\sqrt {x^2+y^2+z^2} \quad \quad \quad \quad \quad \nonumber \\ 
f\left( {r{\kern 1pt},\,t} \right)=\quad\quad\quad\quad\quad\quad\quad\quad\quad\quad\quad\quad\quad\quad\quad\quad\quad\quad\quad\quad\quad\quad\quad\quad\quad\quad\nonumber\\
\left( {6+2\,k^2r^2-k^4r^4} \right)\cos k\left( {r-c\,t} \right)+\left( {k^5r^5\left( {\pi \mathord{\left/ {\vphantom {\pi {2-\mbox{Si}\left( {k\,r} \right)}}} \right. \kern-\nulldelimiterspace} {2-\mbox{Si}\left( {k\,r} \right)}} \right)-6} \right)\cos c\,k\,t \nonumber \\ 
+k\,r\,\left( {6-k^2r^2} \right)\,\sin k\left( {r-c\,t} \right)+k^5r^5\,\mbox{Ci}\left( {k\,r} \right)\,\sin c\,k\,t \quad \quad \quad \quad \quad \quad \quad 
\end{eqnarray}
It is obvious that (\ref{eq28}), (\ref{eq21}), and (\ref{eq19}) inserted into (\ref{eq5}) do not yield the electric field (\ref{eq16}) as given by the Li\'{e}nard-Wiechert solution, since the sine and cosine integrals do not occur in (\ref{eq16}) when (\ref{eq27}) is inserted. Furthermore, insertion of (\ref{eq28}) and (\ref{eq21}) into (\ref{eq5}) does not yield the magnetic field as given by (\ref{eq16}), since the rotation of (\ref{eq21}) results already in the magnetic field (\ref{eq16}), but the rotation of (\ref{eq28}) does not vanish. 

These conclusions were already drawn in Sect. 3 of Ref. [2], but they could not be based on an explicit solution like (\ref{eq28}). At this point the deeper reason for the discrepancy is still not clear, but the puzzle will be solved in the next Section when we take a closer look at the inhomogeneous wave equation.

\vskip 30pt
\section{Attempt to solve an inhomogeneous wave equation}
In Sect. 6 of Ref. 2 we have used a method of solution applied to (\ref{eq7}) which resulted in a potential at variance with (\ref{eq10}). Here we analyze the wave equation for the magnetic field which results from elimination of the electric field from (\ref{eq2}) and (\ref{eq4}):
\begin{equation}
\label{eq29}
\Delta \vec {B}-\frac{1}{c^2}\frac{\partial ^2\vec {B}}{\partial t^2}=-\frac{4\pi }{c}\,\mbox{rot}\,\vec {j}
\end{equation}
In this linear equation one may split the magnetic field into two contributions
\begin{equation}
\label{eq30}
\vec {B}=\vec {B}_0 +\vec {B}_1 
\end{equation}
where the first one obeys the Poisson equation
\begin{equation}
\label{eq31}
\Delta \vec {B}_0 =-\frac{4\pi }{c}\,\mbox{rot}\,\vec {j}
\end{equation}
Its solution enters as a source into a wave equation for the second contribution:
\begin{equation}
\label{eq32}
\Delta \vec {B}_1 -\frac{1}{c^2}\frac{\partial ^2\vec {B}_1 }{\partial t^2}=\frac{1}{c^2}\frac{\partial ^2\vec {B}_0 }{\partial t^2}
\end{equation}
Choosing the Hertzian dipole $\dot {\vec {p}}\left( t \right)$ as the source in (\ref{eq31}) we find
\begin{equation}
\label{eq33}
\vec {B}_0 =\frac{\dot {\vec {p}}\left( t \right)\times \vec {r}}{c\,r^3}
\end{equation}
When this is substituted into (\ref{eq32}), all textbooks (e.g. [4] and [5]) suggest as a solution the retarded volume integral:
\begin{equation}
\label{eq34}
\vec {B}_1 \left( {\vec {x}{\kern 1pt},\,t} \right)=-\frac{1}{4\pi \,c^3}\int\!\!\!\int\!\!\!\int {\frac{{\ddot{\dot{\vec {p}}}}\left( {t-{r'} \mathord{\left/ {\vphantom {{r'} c}} \right. \kern-\nulldelimiterspace} c} \right)\times \vec {r}\,'}{r'{\kern 1pt}^3}} \frac{d^3r'}{\left| {\vec {x}-\vec {r}\,'} \right|}
\end{equation}
Due to the axial symmetry of the problem it is sufficient to calculate the $y$-component of (\ref{eq34}) at $y=0$:
\begin{equation}
\label{eq35}
B_{1y} =-\frac{1}{4\pi \,c^3}\int\limits_0^\infty {dr'} \int\limits_0^\pi {d\theta '\int\limits_0^{2\pi } {d\varphi '\,\frac{\ddot{\dot{p}}\left( {t-{r'} \mathord{\left/ {\vphantom {{r'} c}} \right. \kern-\nulldelimiterspace} c} \right)\,\,\cos \varphi '\sin ^2\theta '\,\,}{R } } } 
\end{equation}
where $R$ is defined in (\ref{eq24}). Inserting (\ref{eq27}) and performing the integration over the angles yields:
\begin{equation}
\label{eq36}
B_{1y} =\frac{\omega ^3\sin \theta }{3\,c^3}\left[ {\int\limits_0^r {\frac{r'\cos \omega \left( {t-{r'} \mathord{\left/ {\vphantom {{r'} c}} \right. \kern-\nulldelimiterspace} c} \right)}{r^2}\,dr'\,+\,\int\limits_r^\infty {\frac{r\cos \omega \left( {t-{r'} \mathord{\left/ {\vphantom {{r'} c}} \right. \kern-\nulldelimiterspace} c} \right)}{r'^2}\,dr'} } } \right]\quad\quad
\end{equation}
Finally, with $\omega =c\,k$ one obtains for the Cartesian components of $\vec {B}_1 $:
\begin{eqnarray}
\label{eq37}
\vec {B}_1 \left( {\vec {x},\,t} \right)=\frac{k\,f\left( {r,\,t} \right)}{3\,r^3}\frac{\vec {p}\times \vec {r}}{p}  \quad \quad \quad \quad \quad \quad \quad \quad \quad \quad \\ 
 f\left( {r,\,t} \right)=\left( {1+k^2r^2} \right)\cos k\left( {r-c\,t} \right)-\left( {1+k^3r^3\left( {\pi \mathord{\left/ {\vphantom {\pi {2-}}} \right. \kern-\nulldelimiterspace} {2-}\mbox{Si}\left( {k\,r} \right)} \right)} \right)\cos c\,k\,t \nonumber\\ 
+k\,r\sin k\left( {r-c\,t} \right)-k^3r^3\,\mbox{Ci}\left( {k\,r} \right)\sin c\,k\,t \quad \quad \quad \quad \quad \quad \quad \quad \quad
\nonumber
\end{eqnarray}
It is obvious that (\ref{eq33}) and (\ref{eq37}) inserted into (\ref{eq30}) do not yield the Li\'{e}nard-Wiechert magnetic field (\ref{eq16}). As in Sect. 3 we encounter again an ambiguity of solutions.

Furthermore, the Laplace operator applied on (\ref{eq37}) yields:
\begin{equation}
\label{eq38}
\Delta {\kern 1pt}\vec {B}_1 \left( {\vec {x}{\kern 1pt},\,t} \right)=-\frac{k^3}{\,r^3}\,\cos k\left( {r-c\,t} \right)\frac{\vec {p}\times \vec {r}}{p}=\frac{1}{c^2}\frac{\partial ^2\vec {B}_0 \left( {\vec {x}{\kern 1pt},\,t-r \mathord{\left/ {\vphantom {r c}} \right. \kern-\nulldelimiterspace} c} \right)}{\partial t^2}
\end{equation}
Substituting this into the wave equation (\ref{eq32}) and integrating twice over time we obtain:
\begin{equation}
\label{eq39}
\vec {B}_0 \left( {\vec {x},\,t-r/c} \right)-\vec {B}_1 \left( {\vec {x},\,t} \right)=\vec {B}_0 \left( {\vec {x},\,t} \right)
\end{equation}
This relationship is certainly not satisfied by the insertion of (\ref{eq37}). Surprisingly, it turns out that the ``solution'' (\ref{eq37}) does not satisfy the wave equation (\ref{eq32}) which it is supposed to solve!

The root of the discrepancies identified in Ref. [2] is now revealed: It is wrong to assume that the retarded (or advanced) functions would solve inhomogeneous wave equations like (\ref{eq32}). In case of the potential equations (7, 8) the deficiency of (10, 11) was overlooked, since the chosen point sources did not explicitly enter into the differential equations as already remarked at the end of Sect. 2. If one chooses, however, a spatially extended source like (\ref{eq33}) the inability of the retarded integral (\ref{eq34}) to solve the pertinent wave equation surfaces straight away.

It should be noted that the problem encountered is not caused by the splitting (\ref{eq30}). For the formal solution of (\ref{eq32}) any arbitrary function could be inserted into the inhomogeneity on the right-hand-side, quite independently of the special choice (\ref{eq33}) imposed by the constraints of Maxwell's equations.

\vskip 30pt
\section{Discussion and conclusion}
One may wonder how the mistake as to introduce retarded solutions into the framework of classical electrodynamics came about. It is certainly not due to Maxwell himself who did not deal with inhomogeneous wave equations at the price of committing a mistake (see Ref. [2] and [6]). It appears that Duhamel's principle, namely to construct solutions of the inhomogeneous equations out of solutions for the homogeneous equations, has been wrongly applied to the wave equation. Originally the method was applied to the diffusion equation ${\partial T} \mathord{\left/ {\vphantom {{\partial T} {\partial t}}} \right. \kern-\nulldelimiterspace} {\partial t}=D\,\Delta T+q\left( {\vec {x}{\kern 1pt},\,t} \right)$ which has a built-in property: Any change in the source has an instantaneous influence on the temperature in the whole space. This may be an unphysical assumption, but it is built into the transport equation and allows Duhamel's principle to be used. There is a common time parameter which describes consistently changes in the heat production and the instant reaction of the temperature field.

The wave equation, however, describes travelling fields which are disconnected from their source the more the longer they travel. It is therefore not reasonable to construct solutions of the inhomogeneous equations with solutions that have ``forgotten'' their source. As pointed out in the Introduction and in [2] it makes no sense to connect the fields and their -- possibly already extinguished -- sources at the same time. Equation (\ref{eq39}) reveals nicely this logical inconsistency, as this relationship constitutes an impossible connection between the field $\vec {B}_0 $ at the same place now and in the past. 

The question remains whether there are other methods to solve Maxwell's equations. Our position in this respect is the same as Dunn's [3]. The instantaneous quasi static fields may be calculated from elliptic equations such as (\ref{eq18}) or (\ref{eq31}). In case of open circuits one may complement (\ref{eq31}) by a contribution containing the gradient of the Coulomb field. The induction law (\ref{eq2}) allows determining also an instantaneous rotational electric field from a given instantaneous magnetic field (see Appendix). 

The travelling wave fields, however, follow from hyperbolic equations which must be kept separate from the elliptic equations. This cannot be done without committing inconsistencies which became obvious already in Maxwell's Treatise. As long as one is not prepared, however, to modify Maxwell's system, this consequence is inescapable. Fortunately it is possible to solve the homogeneous hyperbolic wave equations with Cauchy-type boundary conditions that contain a physics different from Maxwell's equations such as Ohm's law, for example. A closed theory comprising both the phenomena of instantaneous quasistatic fields and travelling wave fields is not available yet.

\newpage
\vskip 30pt 
\noindent 
\textbf{Appendix}

\noindent
As Donald Dunn [3] pointed out, the technically important fields which occur in condensers, magnets, transformers etc. can be calculated for a given charge and current distribution from the integrals: 
\setcounter{equation}{0} 
\renewcommand{\theequation}{A.\arabic{equation}} 
\begin{equation}
\vec {E}_C \left( {\vec {x}{\kern 1pt},\,t} \right)=\int\!\!\!\int\!\!\!\int \rho \left( {\vec {x}\,'{\kern 1pt},\,t} \right)\frac{\vec {x}-\vec {x}\,'}{\left| {\vec {x}-\vec {x}\,'} \right|^3}\,d^3x' 
\end{equation}
\begin{equation}\vec {E}_i \left( {\vec {x}{\kern 1pt},\,t} \right)=\frac{-1}{4\pi \,c}\int\!\!\!\int\!\!\!\int {\frac{\partial \vec {B}_i \left( {\vec {x}\,'{\kern 1pt},\,t} \right)}{\partial t}} \times \frac{\vec {x}-\vec {x}\,'}{\left| {\vec {x}-\vec {x}\,'} \right|^3}\,d^3x'
\end{equation} 
\begin{equation}
\vec {B}_i \left( {\vec {x}{\kern 1pt},\,t} \right)=\frac{1}{c}\int\!\!\!\int\!\!\!\int {\left( {\vec {j}\left( {\vec {x}\,'{\kern 1pt},\,t} \right)+\frac{1}{4\pi }\frac{\partial \vec {E}_C \left( {\vec {x}\,'{\kern 1pt},\,t} \right)}{\partial t}} \right)} \times \frac{\vec {x}-\vec {x}\,'}{\left| {\vec {x}-\vec {x}\,'} \right|^3}\,d^3x'
\end{equation} 
\noindent 
These instantaneous fields fall off with $1/r^3$ and appear as sources in the wave equations:
\begin{equation}\Delta \vec {E}_W -\frac{1}{c^2}\frac{\partial ^2\vec {E}_W }{\partial t^2}=\frac{1}{c^2}\frac{\partial ^2\vec {E}_i }{\partial t^2}
\end{equation}
\begin{equation}
\Delta \vec {B}_W -\frac{1}{c^2}\frac{\partial ^2\vec {B}_W }{\partial t^2}=\frac{1}{c^2}\frac{\partial ^2\vec {B}_i }{\partial t^2}
\end{equation}
\noindent
The set of equations (A.1) -- (A.5) is entirely equivalent to Maxwell's equations (1) -- (4), if we substitute 
\begin{equation}
\vec {E}=\vec {E}_C +\vec {E}_i +\vec {E}_{W\;,\quad } \vec {B}=\vec {B}_i +\vec {B}_W
\end{equation}

The wave equations (A.4), (A.5), however, cannot be used close to the sources, since the same time appears at the observation point and at the source. Following Maxwell himself one must drop the right-hand-sides of (A.4) and (A.5) so that one has only to solve homogeneous wave equations under suitable initial and boundary conditions.

The set of equations (A.1) -- (A.3) remains still useful to describe the instantaneous fields close to the sources $\rho \,\,\mbox{and}\,\vec {j}$. In practice this separation of the ``near field'' and the ``far field'' world works quite well, but from a theoretical point of view a connection between the two kinds of fields would be desirable. 

\newpage
\vskip 30pt
\begin{eref}
\bibitem{Onoochin} V. V. Onoochin, \textit{On non-equivalence of Lorentz and Coulomb gauges within classical electrodynamics}, Annales de la Fondation Louis de Broglie, \textbf{27}, 163, (2002). 
\bibitem{Engelhardt} W. Engelhardt, \textit{Gauge invariance in classical electrodynamics,} Annales de la Fondation Louis de Broglie, \textbf{30}, 157 (2005).
\bibitem{Dunn} D. A. Dunn, \textit{Models of Particles and Moving Media}, Academic Press, New York and London (1971), page 118ff.
\bibitem{Jackson} J. D. Jackson, \textit{Classical Electrodynamics, }Second Edition, John Wiley {\&} Sons, Inc., 
New York (1975), Chapter 6.
\bibitem{Becker} R. Becker, F. Sauter, \textit{Theorie der Elektrizit\"{a}t}, B. G. Teubner Verlagsgesellschaft, Stuttgart (1962), {\S}67.
\bibitem{Maxwell} J. C. Maxwell, \textit{A Treatise on Electricity and Magnetism}, 
Dover Publications, Inc., New York (1954), Vol. 2, Articles 783, 784. 
\end{eref}

\man{3 avril 2011}
\end{document}